\documentclass{moriond}

\def\be{\begin{equation}}
\def\ee{\end{equation}}
\def\bea{\begin{eqnarray}}
\def\eea{\end{eqnarray}}

\usepackage{xspace}
\usepackage[utf8]{inputenc} 

\def\fsfd{\ensuremath{f_s / f_d}\xspace}
\def\runI{Run~1\xspace}
\def\runII{Run~2\xspace}
\def\CL{\ensuremath{(95\%~\texttt{CL})}\xspace}

\def\tev{\ensuremath{\mathrm{\,Te\kern -0.1em V}}\xspace}
\def\mevcc{\ensuremath{{\mathrm{\,Me\kern -0.1em V\!/}c^2}}\xspace}
\def\gevcc{\ensuremath{{\mathrm{\,Ge\kern -0.1em V\!/}c^2}}\xspace}
\def\invfb   {\ensuremath{\mbox{\,fb}^{-1}}\xspace}
\def\ps   {\ensuremath{{\rm \,ps}}\xspace}
\def\sqs   {\ensuremath{\protect\sqrt{s}}\xspace}

\def\bmm{\ensuremath{B^0_{(s)} \rightarrow \mu^+ \mu^-}\xspace}
\def\bsmm{\ensuremath{B^0_s \rightarrow \mu^+ \mu^-}\xspace}
\def\bdmm{\ensuremath{B^0 \rightarrow \mu^+ \mu^-}\xspace}

\newcommand{\Photo}{\includegraphics[height=35mm]{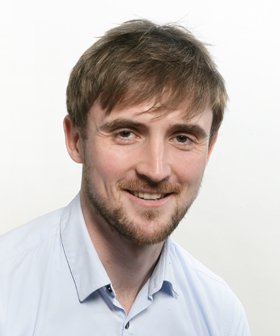}}

\begin{document}
\vspace*{4cm}
\title{Rare B and strange decays}

\author{ S. Tolk \\ on behalf of the LHCb collaboration. }
\address{ University of Cambridge, Cavendish Laboratory, \\ JJ Thomson Ave, Cambridge CB3 0HE, UK }

\maketitle\abstracts{ 
Several deviations from the Standard Model predictions have been recently observed in the decays mediated by $b \rightarrow s l^+ l^-$ transitions. 
These could be pointing towards new vector-current contributions or could be explained by underestimated charm-loop effects. 
New results from an LHCb \runI $B^+\rightarrow K^+ \mu^+ \mu^- $ analysis that includes the decays via intermediate charm-resonances are discussed.
Also, new results from the fully leptonic rare modes searches are presented. This includes the latest \runI and \runII \bmm analysis from LHCb where the \bsmm candidates are used to determine the effective lifetime of the \bsmm decays - 
a pioneering result that in the future will solve the current ambiguity in the (pseudo-)scalar contributions.}

\section{Introduction}

Flavour changing up-up or down-down type quark transitions are rare in the Standard Model (SM). Apart from being forbidden at the tree level, they are further suppressed 
at the loop level by the off-diagonal CKM elements, GIM suppression and for di-leptonic decays also helicity suppression. Although it is experimentally challenging 
to separate the rare modes from large SM background, the observables in these rare processes are sensitive to New Physics (NP) effects far beyond the energies directly 
accessible in colliders.

Rare decays can be described in an effective field theory. Using an effective Hamiltonian the $B$ decay amplitude can be schematically written as
\be
    A(B\rightarrow f) = \langle f | \mathcal{H}_{eff} | B \rangle = \frac{G_F}{\sqrt{2}} \sum_i  \lambda_{CKM}  C_i(\mu_b)  \langle f | \mathcal{Q}_{i}(\mu_b) | B \rangle,
\ee
where $G_F$ is the weak coupling constant, $\lambda_{CKM}$ are the CKM elements, $C_i$ are the Wilson coefficients containing perturbative short-distance effects (evaluated at 
the energy scale $\mu_b$) and $Q_i$ denote the operators containing the non-perturbative and long-distance effects.
Three types of operators are relevant for the $B$ decays discussed here: the electromagnetic penguin operator ($Q^{(')}_7$), the vector and axial-vector semileptonic 
operators ($Q^{(')}_9$ and $Q^{(')}_{10}$), the scalar and pseudo-scalar operators ($Q^{(')}_S$ and $Q^{(')}_{P}$).
NP effects can alter the corresponding Wilson coefficients. The coefficient values are determined from global analyses that include experimental results from the $b\rightarrow s l^+l^-(\gamma)$ transition processes.

The most recent global $b\rightarrow s l^+l^-(\gamma)$ analyses~\cite{Altmannshofer:2014rta,Hurth:2016fbr,Descotes-Genon:2015uva} agree that there are tensions
between the Wilson coefficients preferred by the data and the values predicted in the SM. The tension is driven by the measured $B^0\rightarrow K^{*}\mu^+ \mu^-$ angular 
observables~\cite{Aaij:2015oid} and several differential branching fractions of $b\rightarrow s l^+l^-$ type decays~\cite{Aaij:2015oid,Aaij:2014pli} 
which tend to be lower than their SM predictions~\footnote{The new ATLAS and CMS $B^0 \rightarrow K^{0*}\mu^+\mu^-$ angular analysis results were presented at Moriond EW 2017. In case of the simplest vector-current NP scenario 
in $C_9$, the latest CMS result decreases and the ATLAS result increases the deviation from the SM value. 
If both results are included in the global analysis then the results are consistent with the previous picture and the tension remains strong~\cite{Altmannshofer:2017fio,Ahmed:2017vsr,Blake:2017wjz}.}. 
These tensions could be explained by short-distance contributions from new particles or indicate a problem with the SM hadronic contributions predictions.

\section{Resonance effects in the vector current ($C_9$)}\label{subsec:prod}

One way to explain the tension within the SM is to allow for sizeable long-distance effects in di-muon mass regions far from the pole masses of the resonances.
LHCb studies this possibility in a $B^+\rightarrow K^+ \mu^+ \mu^-$ analysis that includes the resonances and measures the relative phases of the short-distance and 
the narrow-resonance amplitudes in a wide di-muon mass spectrum.

The results are determined from a fit to the CP-averaged differential decay rate of $B^+ \to K^+ \mu^+ \mu^-$ decays in \runI LHCb data:
\bea
    \frac{d \Gamma}{dq^2} =
    &=& \frac{G^2_F \alpha^2 |V_{tb}V^*_{ts}|^2}{128\pi^2} |k| \beta 
        \Bigg\{ 
            \frac{2}{3} |k|^2 \beta^2 | C_{10} f_+ (q^2)|^2 + \frac{4 m^2_{\mu}(m^2_B-m^2_K)^2}{q^2 m^2_B} |C_{10} f_0 (q^2)|^2 \nonumber \\
    & &+ |k|^2 \bigg[ 1-\frac{1}{3} \beta^2 \bigg] \bigg| C_9 f_+ (q^2) + 2 C_7 \frac{m_b+m_s}{m_B + m_K} f_T(q^2) \bigg|^2 
        \Bigg\} 
\eea
where $k$ is the kaon momentum in the $B^+$ meson rest frame, $m_K$, $m_B$, $m_s$, $m_b$ and $m_{\mu}$ are the respective particle masses,
$\beta^2 = 1- 4m^2_{\mu} / q^2$ and the constants $G_F$, $\alpha$, and $V_{tq}$ are the Fermi constant, the QED fine structure constant
and CKM matrix elements. The $f_{0,+,T}$ are the scalar, vector and tensor  $B\rightarrow K$ form factors.
The Wilson coefficient $C_7$ is small and fixed to the SM value. The coefficient $C_9$ is redefined to include the long-distance effects from the hadronic resonances:
\be
    C^{eff}_9 = C^9 + \sum_j \eta_j e^{i\delta_j} A^{res}_j(q^2).
\ee
The di-muon mass distributions of the resonances, $A^{res}$, are modelled with a Flatt\'e function for $\psi(3770)$, and with Breit-Wigner functions for
the $\omega$, $\rho^0$, $\phi$, $J/\psi$, $\psi(2S)$, $\psi(4040)$, $\psi(4160)$, and $\psi(4415)$ resonances.
The widths of all the resonances and the pole masses for all except the $J/\psi$ and the $\psi(2S)$ are fixed to their known values. 
Contributions from other broad resonances and the hadronic continuum states are small and ignored at this point. Instead, the short-distance contribution is normalised 
to the $B^+~\rightarrow~J/\psi(\rightarrow \mu^+ \mu^-) K^+$ decays and the magnitudes of ($\eta_j$) and the relative phases ($\delta_j$) between the resonances and the short-distance 
contribution are allowed to vary in the fit.

\begin{figure}
\begin{minipage}{\linewidth}
\centerline{\includegraphics[trim=0 14.5cm 0 0 ,clip, width=\linewidth]{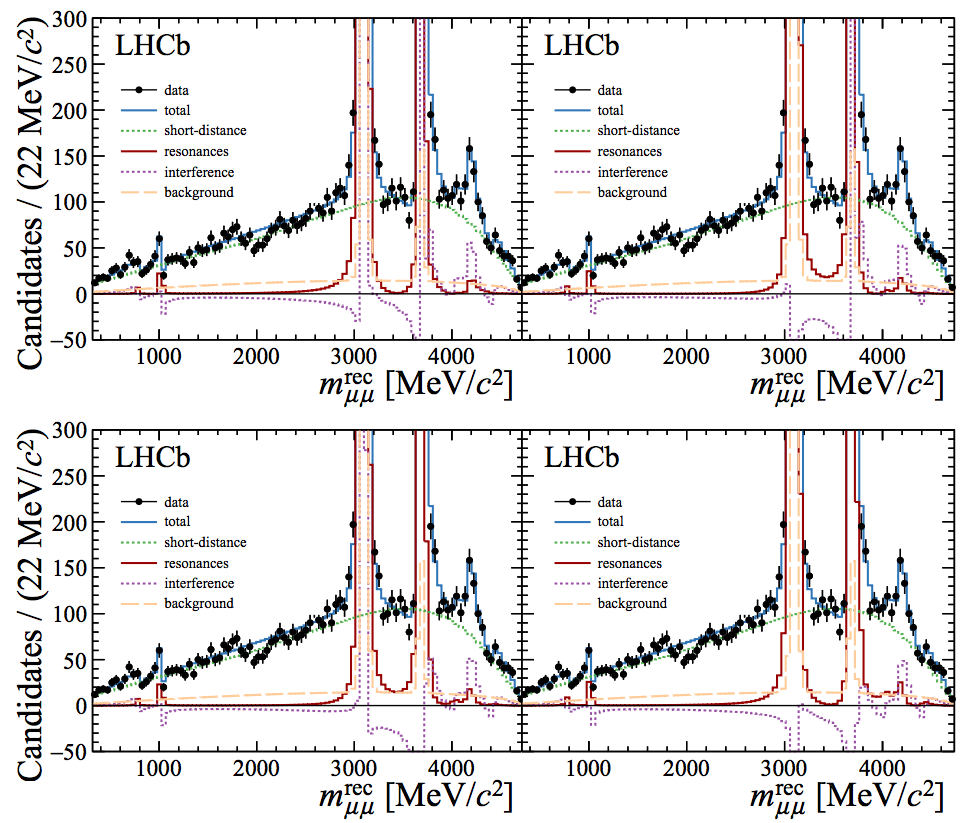}}
\end{minipage}
\caption{Fits to the di-muon invariant mass distribution of LHCb \runI data. The fit has four solutions, depending on the relative phases of the two most dominant resonances:
$J/\psi$ and $\psi(2S)$. The fit with negative (positive) $J/\psi$ and negative (negative) $\psi(2S)$ phases is shown on the right (left). The interference component denotes 
the interference between the short and long-distance contributions. The details are given in the paper~\protect\cite{Aaij:2016cbx}.}
\label{fig:kmm}
\end{figure}

The fit to the di-muon mass spectrum is shown in Figure~\ref{fig:kmm}. Four solutions arise due to the ambiguities in the signs of the $J/\psi$ and $\psi(2S)$ phases.
The values of the $J/\psi$ phases are compatible with $\pm \frac{\pi}{2}$, which means the interference with the short-distance component in
di-muon mass regions far from the resonances is small. The measurement of the Wilson coefficients $C_9$ and $C_{10}$ prefers $|C_9|>|C^{SM}_{9}|$ and $|C_{10}|<|C^{SM}_{10}|$.
If $C_{10}$ is constrained to its SM value, then the fit prefers $|C_9|<|C_9^{SM}|$ which is in agreement with the global analysis~\cite{Aaij:2016cbx}.
The branching fraction for the short-distance component alone is:
\be
    \mathcal{B}(B^+\rightarrow K^+ \mu^+ \mu^- )  = (4.37 \pm 0.15 (\texttt{stat}) \pm 0.23 (\texttt{syst})) \times 10^{-7},
    \label{eq:short_bf}
\ee
which is in good agreement with the previous result from the exclusive analysis~\cite{Aaij:2014pli}. Note that unlike the previous measurement, 
the branching fraction in Equation~\ref{eq:short_bf} does not rely on extrapolation over the excluded $q^2$ regions.

\section{The very rare decays \bmm}

The scalar ($C_S$) and pseudo-scalar ($C_P$) Wilson coefficients can be determined in the fully leptonic $B$ decays. In the SM these modes are dominated by the helicity suppressed axial-vector 
current ($C_{10}$) contributions and the branching fractions are very precisely predicted~\cite{Bobeth:2013uxa}. 
The latest parametric input values~\footnote{Mostly the $B^0_s$ lifetime, the relative $B^0_s$ decay width difference ($\frac{\Delta\Gamma_s}{2\Gamma_s}$), $B^0_s$
decay constant ($f_{B_s}$) and the CKM elements $|V_{tb}|$ and $|V_{ts}|$.} reduce the relative uncertainty on the \bsmm branching fraction to $4.5\%$~\protect\cite{Fleischer:2017ltw}:
\be
 \mathcal{B}(\bsmm) = (3.57 \pm 0.16) \times 10^{-9}.
\ee
Although negligible in the SM, the (pseudo-)scalar contributions are free of the helicity suppression. The di-leptonic decays of $B$ mesons are therefore particularly sensitive to 
new (pseudo-)scalars~\footnote{Under the right conditions, the decays can be sensitive to new vector bosons such as $Z^{\prime}$ with masses up to $160\tev$ and to new scalars with masses up 
to $1000\tev$~\cite{Buras:2014zga}.}.

The \bsmm and \bdmm decays have been experimentally searched for since 1985. The \bsmm decays were finally observed in the combined CMS and LHCb Run 1 analysis~\cite{CMS:2014xfa}. The measured \bsmm branching 
fraction was lower than but compatible with the SM prediction. In combination with the unexpectedly high \bdmm candidate yield, the relative \bdmm and \bsmm branching fraction ratio deviated 
from the precise SM expectation~\footnote{Or any other Minimal Flavour Violating model prediction.} by $2.3\sigma$. 

The measured \bsmm significance in the \runI ATLAS data remains below the evidence level ($2\sigma$)~\cite{Aaboud:2016ire}. The \bdmm yield in ATLAS data is compatible with the background expectations and 
ATLAS sets an upper limit on the \bdmm branching fraction at $4.2\times 10^{-10} ~ \CL$. Given the large uncertainties, the results are in agreement with both the SM predictions and the combined CMS and LHCb results.

The most recent \bmm results are from LHCb and include proton-proton collision data from \runII: $\mathcal{L}=0.3~\invfb$ from 2015 and $\mathcal{L}=1.1~\invfb$ from 2016. Both \runII samples were recorded at $\sqs = 13\tev$. 
This time the selection is optimised for the \bdmm mode and the signal detection efficiencies are estimated individually for each mode in order to account for the small differences. Several key steps 
of the analysis are significantly improved with respect to the \runI analysis~\cite{Aaij:2013aka}: the rejection power of the most dangerous background contributions from the doubly mis-identified $B^0_{(s)}\rightarrow h^+ h^{(')-}$ 
modes is increased by $50\%$ at the expense of only $10\%$ of the signal loss; the multivariate Boosted Decision Tree classifier that separates the true two-body decays from the random combinations now includes new Boosted Decision Tree based muon isolation variables; the background estimates for $B^0_{(s)}\rightarrow h^+ h^{'-}$, $B^0 \rightarrow \pi^- \mu^+ \nu_{\mu}$, and $B^0_s \rightarrow K^- \mu^+ \nu_{\mu}$ 
are validated by fits to the $\pi^- \mu^+$ and $K^- \mu^+$ invariant mass spectra in data after correcting for the hadron-to-muon mis-identification probabilities.

The relative $B^0_s$ and $B^{0(+)}$ meson production fraction (\fsfd) value measured by the LHCb on the \runI data~\cite{LHCb-CONF-2013-011} is used in normalising the signal branching fractions in both runs.
The \fsfd at the higher proton-proton collision energy in \runII is determined by comparing the ratio of the efficiency corrected $B^0_s \rightarrow J/\psi \phi$ and $B^+ \rightarrow J/\psi K^+$ yields in \runI and \runII data.
The relative yields are stable and the \runII-to-\runI ratio is included in the \runII normalisation as a constraint from an auxiliary measurement to account for the uncertainty.

\begin{figure}
\begin{minipage}{0.49\linewidth}
\centerline{\includegraphics[width=1.0\linewidth]{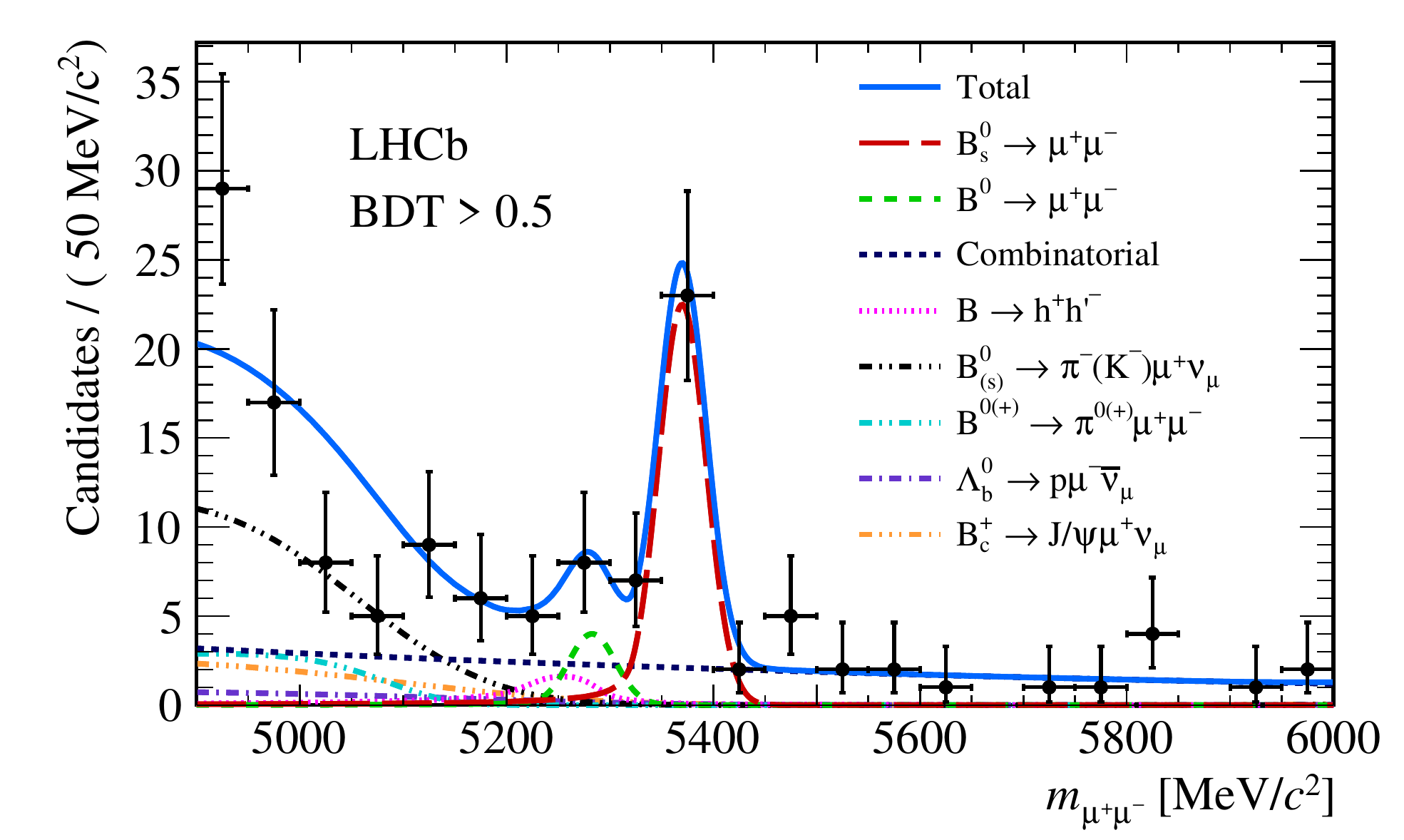}}
\end{minipage}
\hfill
\begin{minipage}{0.49\linewidth}
\centerline{\includegraphics[width=1.0\linewidth]{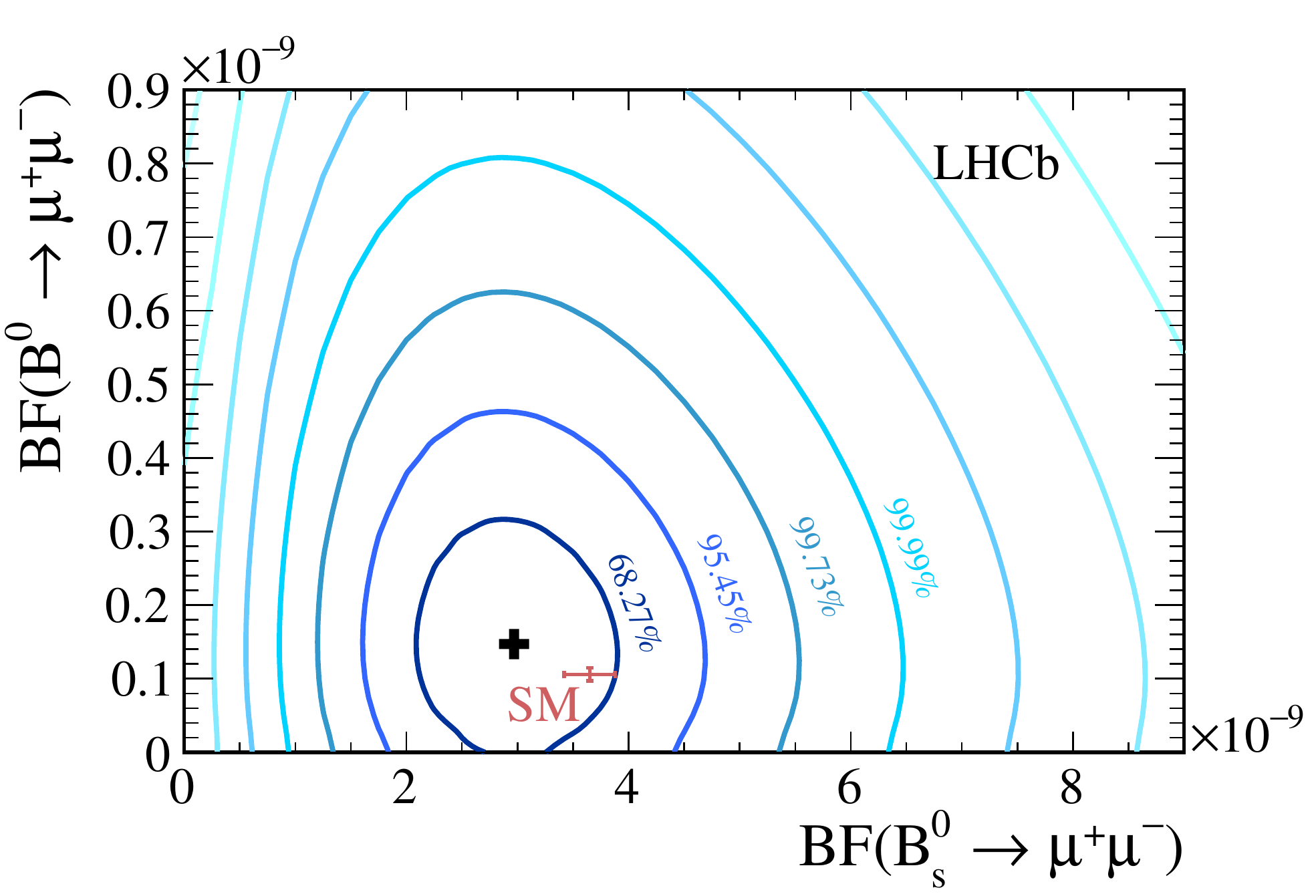}}
\end{minipage}
\caption{Unbinned maximum likelihood fit of the di-muon invariant mass distribution, shown for the best \bmm candidates in \runI and \runII data ($BDT>0.55$, left). 
Profile-likelihood scan and the resulting likelihood contours for the \bmm branching fractions (left). The Standard Model expectation is shown in red~\protect\cite{Aaij:2017vad}.}
\label{fig:bmm}
\end{figure}

The most recent LHCb \bmm results using the full \runI data sample and $\mathcal{L}~=~1.4\invfb$ of \runII data are~\cite{Aaij:2017vad}:
\bea
  \mathcal{B}(\bsmm) & = & (3.0 \pm 0.6^{+0.3}_{-0.2}) \times 10^{-9}     ~ (7.8\sigma), \\
  \mathcal{B}(\bdmm) & < & 3.4  \times 10^{-10}                            ~\CL.
\eea
This is the first observation of the \bsmm decay by a single experiment. 
The measured branching fraction is the most precise result currently available.
The result does not confirm the \bdmm excess seen in the \runI analysis. Overall the agreement between the measured signal branching fractions and the SM predictions has improved (Figure~\ref{fig:bmm}).

\begin{figure}
\begin{minipage}{0.49\linewidth}
    \centerline{\includegraphics[width=1.0\linewidth]{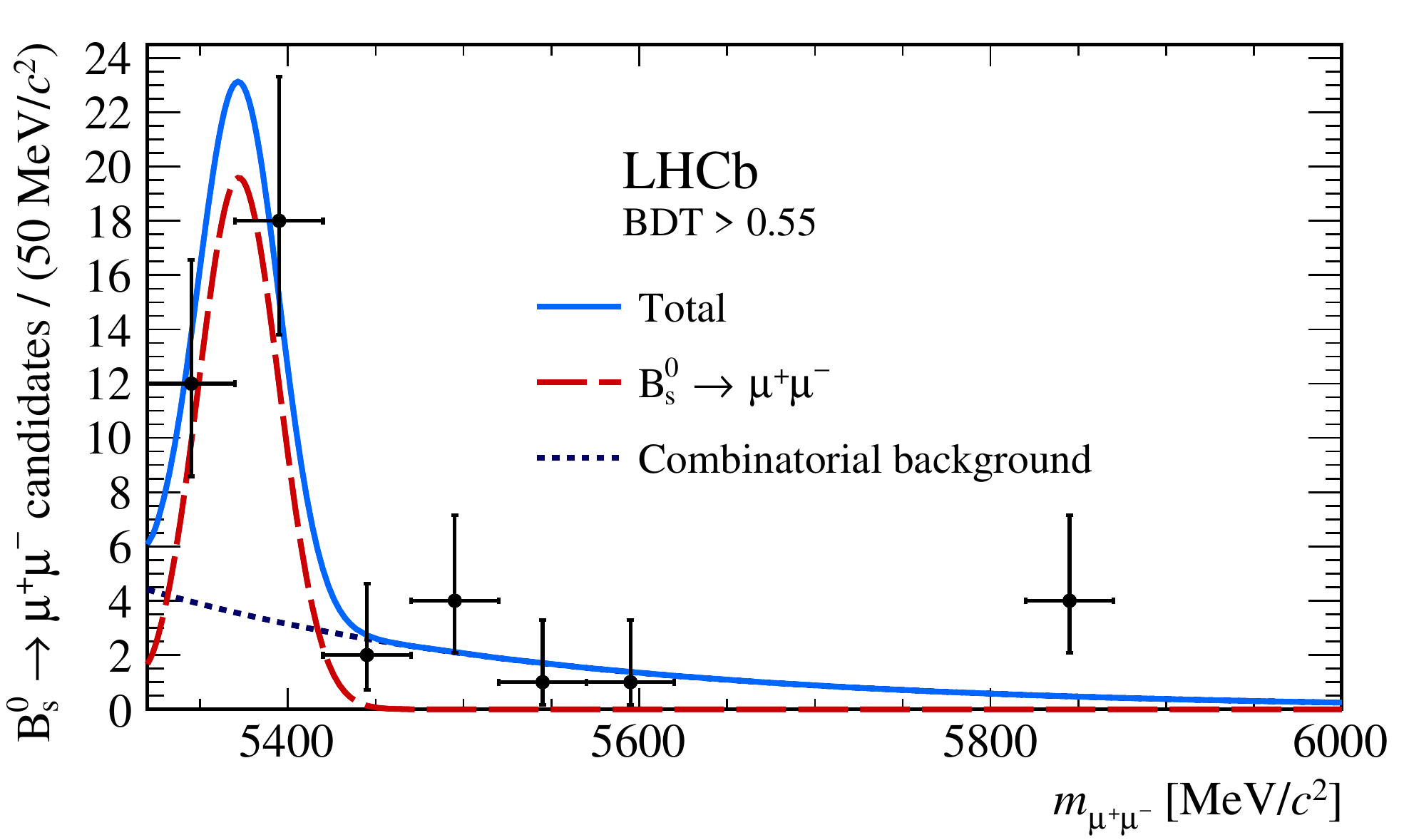}}
\end{minipage}
\hfill
\begin{minipage}{0.49\linewidth}
    \centerline{\includegraphics[width=1.0\linewidth]{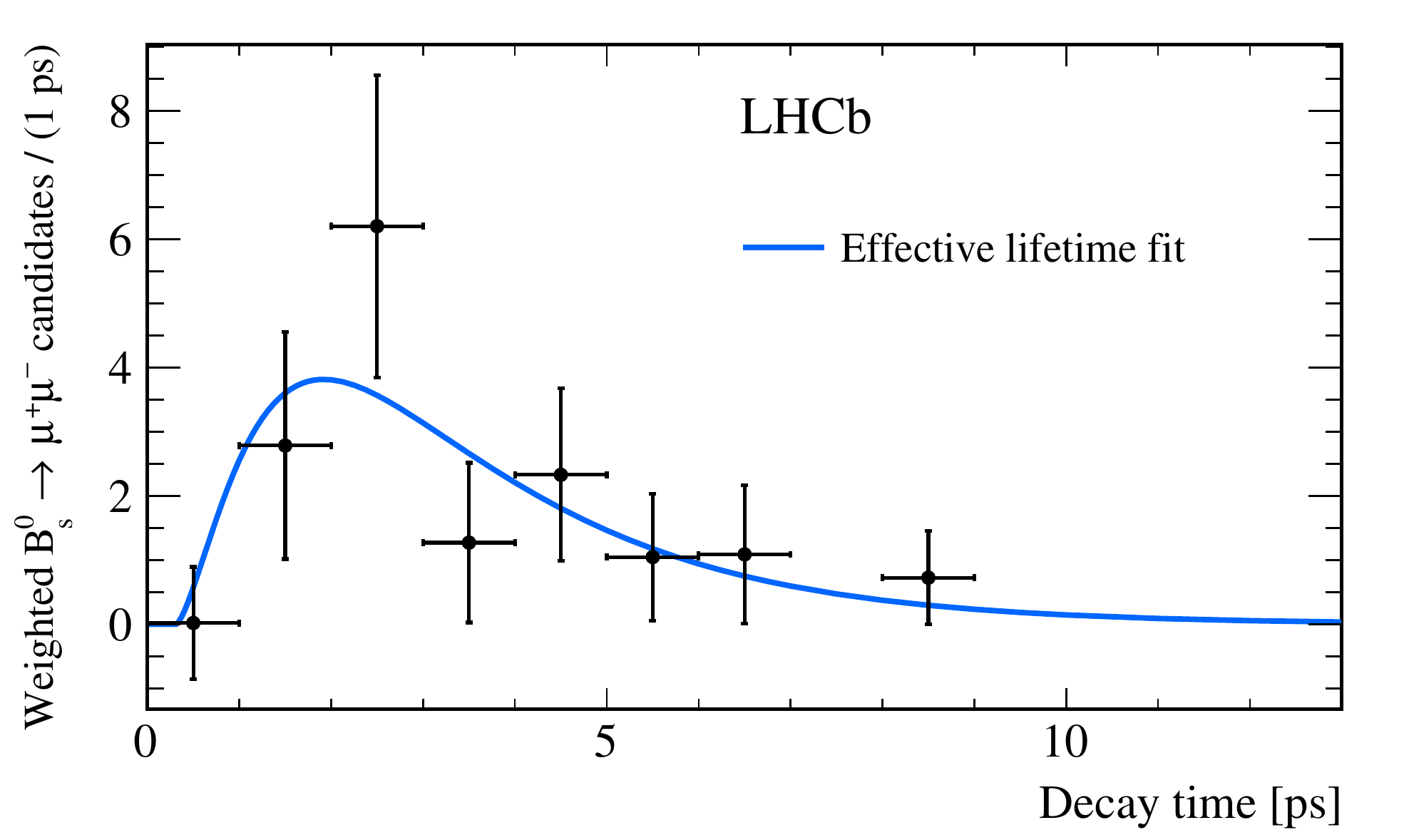}}
\end{minipage}
\caption{Fits relevant for the \bsmm effective lifetime measurement: the di-muon invariant mass distribution fit on the most signal like candidates in \runI and \runII data in higher mass region where only the combinatorial background contribution is significant (left); the lifetime fit on the sWeighted \bsmm signal candidates (right)~\protect\cite{Aaij:2017vad}.}
\label{fig:bmm_t}
\end{figure}

The implications of the new LHCb \bmm results are discussed in several papers~\cite{Altmannshofer:2017wqy,Fleischer:2017ltw,Hussain:2017tdf}. The \bsmm branching fraction 
is found to be especially useful for probing the high multi-\tev mass region of the Two-Higgs-Doublet models in decoupling regime and of the models with leptoquarks. 
In the case of a model independent $C_S = - C_P$ scenario~\footnote{This holds in general for Minimal Flavour Violating New Physics, e.g Minimal Supersymmetric SM.} the current situation leads to two equivalent solutions 
for the (pseudo-)scalar coefficients: one with SM like values and one corresponding to sizeable deviations~\cite{Altmannshofer:2017wqy}. The degeneracy can be solved by measuring the \bsmm \emph{mass-eigenstate-rate-asymmetry}:
\be
 A_{\Delta \Gamma} = \frac{ \Gamma(B^H_s \rightarrow \mu^+ \mu^-)  - \Gamma(B^L_s \rightarrow \mu^+ \mu^-)  } {\Gamma(B^H_s \rightarrow \mu^+ \mu^-)  + \Gamma(B^L_s \rightarrow \mu^+ \mu^-)  },
\ee
which is $+1$ in the SM but could be as low as $-1$ if NP is involved.
The mass-eigenstate-rate-asymmetry can be determined from the \bsmm effective lifetime~\cite{DeBruyn:2012wk}. In the latest analysis LHCb shows that the measurement is possible even with the very limited available statistics. 
The best \bsmm candidates according to the Boosted Decision Tree classifier and the muon identification criteria (Figure~\ref{fig:bmm_t}) are used to determine the \bsmm effective lifetime:
\bea
    \tau(\bsmm) = 2.04 \pm 0.44(\texttt{stat}) \pm 0.05 (\texttt{syst})\ps. 
\eea
The measurement is compatible with the heavy $B$ eigenstate lifetime of $\tau_H=(1.615\pm0.01)\ps$ as expected in the SM.
Due to large statistical uncertainty, no conversion to mass-eigenstate-rate-asymmetry is attempted at this point. 
The result is found to be compatible with $A_{\Delta \Gamma} = +1 (-1)$ at $1.0\sigma (1.4\sigma)$.

\section{$B^0_s\rightarrow \tau^+ \tau^-$ and $K_S \rightarrow \mu^+ \mu^-$ searches}

From all the di-lepton modes, helicity suppression affects $B^0_s\rightarrow \tau^+ \tau^-$ decays the least. The SM branching fraction prediction for the tauonic mode is~\cite{Bobeth:2013uxa} :
\bea    
    \mathcal{B}(B^0_s \rightarrow \tau^+ \tau^-)^{SM} = (7.73\pm0.49)\times 10^{-7},
\eea    
which could be enhanced by a factor of $\sim10^{3}$ in the NP interpretations of the lepton flavour universality anomalies~\cite{Alonso:2015sja,Crivellin:2017zlb}. 
The BaBar collaboration has previously set a limit on the $B^0$ mode~\cite{PhysRevLett.96.241802}: $\mathcal{B}(B^0 \rightarrow \tau^+ \tau^-) < 4.1 \times 10^{-3}~\CL$.
LHCb searches for the $B^0_{(s)} \rightarrow \tau^+ \tau^-$ decays where the tau leptons decay into pions and a neutrino: $\tau^{\pm}\rightarrow \pi^{\pm} \pi^{\mp} \pi^{\pm} \bar{\nu}_{\tau}$. 
The analysis makes use of the intermediate resonance $\rho(770)\rightarrow \pi^+ \pi^-$ to improve the signal selection. The results based on the \runI data lead to the most stringent limits yet~\cite{Aaij:2017xqt}:
\bea
 \mathcal{B} (B^0 \rightarrow \tau^+ \tau^-)  & < & 2.1 \times 10^{-3} ~  \CL , \\
 \mathcal{B} (B^0_s \rightarrow \tau^+ \tau^-)  & < & 6.8 \times 10^{-3} ~  \CL .
\eea
Since neither mode has been experimentally observed, the limits are set by assuming one or the other neutral $B$ meson mode.

Neutral kaon decays to the $\mu^+\mu^-$ final state have been measured for the $K_L$ mass-eigenstate. According to the SM, the $K_S \rightarrow \mu^+ \mu^-$ decays are expected to occur at a very low rate~\cite{Isidori:2003ts}:
$(5.0\pm1.5)\times10^{-12}$.
NP effects (e.g. from light scalars) could raise the SM rate up to the level of $10^{-10}$ while still avoiding constraints from the other measurements. 
The most stringent experimental limit on $K_S \rightarrow \mu^+ \mu^-$ was set by the LHCb analysis on $1~\invfb$ of the \runI data~\cite{Aaij:2012rt} at $\mathcal{B}(K^0_S\rightarrow \mu^+ \mu^-) < 11\times 10^{-9} ~\CL$.
Using its full \runI data sample, LHCb improves the limit~\cite{LHCb-CONF-2016-012}:
\bea
    \mathcal{B}(K_S \rightarrow \mu^+ \mu^- ) &< & 6.9 \times 10^{-9} ~ \CL.
\eea
Note that this is a preliminary limit and is expected to improve after optimising the trigger and selection criteria.

\section{(Pseudo-)scalar-resonance searches}

In 2005 the HyperCP collaboration reported~\cite{PhysRevLett.94.021801} the first evidence for the decay $\Sigma^+ \rightarrow p \mu^+ \mu^-$.
The measured branching fraction:
\bea
    \mathcal{B}(\Sigma^+ \rightarrow p \mu^+ \mu^-) = (8.6^{+6.6}_{-5.4} \pm 5.4)\times 10^{-8},
\eea
is in agreement with the long-distance dominated SM prediction. The di-muon invariant mass of the candidates, however, is clustered around $m_{X^0} = (214.3\pm0.5)\mevcc$. 
The possibility of a (short-lived) di-muon resonance is investigated by the LHCb in several decay modes. 

A direct search in \runI data shows the $ \Sigma^+ \rightarrow p \mu^+ \mu^-$ signal with a significance of $4.0\sigma$~\cite{LHCb-CONF-2016-013}. A scan along the di-muon invariant mass plane shows no evidence for resonances. 
The precision of the branching fraction measurement is expected to be comparable to the precision of the HyperCP result.

Decays of (pseudo-)scalars into muons could also affect the branching fraction of the very rare mode $B^0_{(s)}\rightarrow \mu^+ \mu^- \mu^+ \mu^-$ and significantly enhance the low SM rate~\cite{Dincer:2003zq} ($ \sim 3.5\times 10^{-11} $).
LHCb searches for the four-muon $B$ modes in the full \runI dataset. No signal is found and LHCb considerably improves the existing limits:
\bea
\mathcal{B} (B^0_s \rightarrow \mu^+ \mu^- \mu^+ \mu^- ) &<& 2.5 \times 10^{-9}, \\
\mathcal{B} (B^0   \rightarrow \mu^+ \mu^- \mu^+ \mu^- ) &<& 6.9 \times 10^{-10}, \\
\mathcal{B} (B^0_s \rightarrow S(\mu^+ \mu^-) P(\mu^+ \mu^-) ) &<& 2.2 \times 10^{-9}, \\
\mathcal{B} (B^0 \rightarrow S(\mu^+ \mu^-) P(\mu^+ \mu^-) ) &<& 6.0 \times 10^{-10},
\eea
where all the limits are estimated at 95\% confidence level and for the last two limits $m_S = 2.6\gevcc$ and $m_P = 214.3\mevcc$ are assumed~\cite{Aaij:2236639}.

\section*{Acknowledgments}
The author wishes to acknowledge the financial support from the Herchel Smith Foundation at Cambridge, his LHCb colleagues for helpful comments and thank the Moriond EW 2017 organisers for organising an inspiring conference.

\section*{References}

\bibliography{bib}

\begin{thebibliography}{10}

\bibitem{Altmannshofer:2014rta}
W. Altmannshofer et~al.,
\newblock EPJ C75 (2015) 382, 1411.3161.

\bibitem{Hurth:2016fbr}
T. Hurth et~al.,
\newblock NP B909 (2016) 737, 1603.00865.

\bibitem{Descotes-Genon:2015uva}
S. Descotes-Genon et~al.,
\newblock JHEP 06 (2016) 092, 1510.04239.

\bibitem{Aaij:2015oid}
LHCb, R. Aaij et~al.,
\newblock JHEP 02 (2016) 104, 1512.04442.

\bibitem{Aaij:2014pli}
LHCb, R. Aaij et~al.,
\newblock JHEP 06 (2014) 133, 1403.8044.

\bibitem{Altmannshofer:2017fio}
W. Altmannshofer et~al.,
\newblock (2017), 1703.09189.

\bibitem{Ahmed:2017vsr}
I. Ahmed et~al.,
\newblock (2017), 1703.09627.

\bibitem{Blake:2017wjz}
T. Blake et~al.,
\newblock EPJ Web Conf. 137 (2017) 01001, 1703.10005.

\bibitem{Aaij:2016cbx}
LHCb, R. Aaij et~al.,
\newblock EPJ C77 (2017) 161, 1612.06764.

\bibitem{Bobeth:2013uxa}
C. Bobeth et~al.,
\newblock PRL 112 (2014) 101801, 1311.0903.

\bibitem{Fleischer:2017ltw}
R. Fleischer et~al.,
\newblock (2017), 1703.10160.

\bibitem{Buras:2014zga}
A.J. Buras et~al.,
\newblock JHEP 11 (2014) 121, 1408.0728.

\bibitem{CMS:2014xfa}
LHCb, CMS, V. Khachatryan et~al.,
\newblock Nature 522 (2015) 68, 1411.4413.

\bibitem{Aaboud:2016ire}
ATLAS, M. Aaboud et~al.,
\newblock EPJ C76 (2016) 513, 1604.04263.

\bibitem{Aaij:2013aka}
LHCb, R. Aaij et~al.,
\newblock PRL 111 (2013) 101805, 1307.5024.

\bibitem{LHCb-CONF-2013-011}
LHCb,
\newblock LHCb-CONF-2013-011  (2013).

\bibitem{Aaij:2017vad}
LHCb, R. Aaij et~al.,
\newblock (2017), 1703.05747.

\bibitem{Altmannshofer:2017wqy}
W. Altmannshofer et~al.,
\newblock (2017), 1702.05498.

\bibitem{Hussain:2017tdf}
M. Hussain et~al.,
\newblock (2017), 1703.10845.

\bibitem{DeBruyn:2012wk}
K. De~Bruyn et~al.,
\newblock PRL 109 (2012) 041801, 1204.1737.

\bibitem{Alonso:2015sja}
R. Alonso et~al.,
\newblock JHEP 10 (2015) 184, 1505.05164.

\bibitem{Crivellin:2017zlb}
A. Crivellin et~al.,
\newblock (2017), 1703.09226.

\bibitem{PhysRevLett.96.241802}
BABAR, B. Aubert et~al.,
\newblock PRL 96 (2006) 241802.

\bibitem{Aaij:2017xqt}
LHCb, R. Aaij et~al.,
\newblock (2017), 1703.02508.

\bibitem{Isidori:2003ts}
G. Isidori et~al.,
\newblock JHEP 01 (2004) 009, hep-ph/0311084.

\bibitem{Aaij:2012rt}
LHCb, R. Aaij et~al.,
\newblock JHEP 01 (2013) 090, 1209.4029.

\bibitem{LHCb-CONF-2016-012}
LHCb,
\newblock LHCb-CONF-2016-012  (2016).

\bibitem{PhysRevLett.94.021801}
HyperCP, H.K. Park et~al.,
\newblock PRL 94 (2005) 021801.

\bibitem{LHCb-CONF-2016-013}
LHCb,
\newblock LHCb-CONF-2016-013  (2016).

\bibitem{Dincer:2003zq}
Y. Dincer et~al.,
\newblock PL B556 (2003) 169, hep-ph/0301056.

\bibitem{Aaij:2236639}
LHCb, R. Aaij et~al.,
\newblock JHEP. 1703 (2016) 001. 19 p.

\end{thebibliography}
\end{document}